\documentclass[runningheads,a4paper]{llncs}

\newcommand{\brad}[1]{}
\newcommand{\alex}[1]{}
\newcommand{\sadia}[1]{}
\newcommand{\rekha}[1]{}
\newcommand{\ling}[1]{}
\newcommand{\mct}[1]{}

\newcommand{\concern}[1]{}
\newcommand{\makeappendix}{}

\usepackage[cmex10]{amsmath}
\usepackage{amssymb}
\usepackage{algorithmic}
\usepackage{array}
\usepackage{url}
\usepackage{multirow}
\usepackage{subcaption}
\usepackage{subfig}
\usepackage{cite}
\usepackage{array,multirow}
\usepackage{color}
\usepackage[normalem]{ulem}
\usepackage{booktabs}
\usepackage{tikz}

\usepackage{graphicx}

\makeatletter
\newcommand{\gobblepars}{\@ifnextchar{\par}{\expandafter\gobblepars\@gobble}{}}
\makeatother
\newcommand{\minisec}[1]{\smallskip\noindent\textbf{{#1}.}\ \ \gobblepars}

\newenvironment{tab}[2]
{
\centering
\let\oldarraystretch=\arraystretch
\renewcommand{\arraystretch}{#1}
\setlength{\tabcolsep}{1.25em}
\begin{tabular}{@{}#2@{}}
\toprule
}
{\bottomrule
\end{tabular}
\renewcommand{\arraystretch}{\oldarraystretch}}

\DeclareMathOperator*{\Var}{Var}

\begin{document}

\mainmatter

\title{Reviewer Integration and Performance Measurement for Malware Detection}

\author{
Brad Miller\inst{1}\textsuperscript{\textdagger} \and
Alex Kantchelian\inst{2}\and
Michael Carl Tschantz\inst{3}\and
Sadia Afroz\inst{3}\and \\
Rekha Bachwani\inst{4}\textsuperscript{\textdaggerdbl} \and 
Riyaz Faizullabhoy\inst{2}\and
Ling Huang\inst{5}\and
Vaishaal Shankar\inst{2}\and \\
Tony Wu\inst{2}\and
George Yiu\inst{6}\textsuperscript{\textdagger} \and
Anthony D. Joseph\inst{2}\and
J. D. Tygar\inst{2}
}

\tocauthor{Brad Miller,
Alex Kantchelian,
Michael Carl Tschantz,
Sadia Afroz,
Rekha Bachwani,
Riyaz Faizullabhoy,
Ling Huang,
Vaishaal Shankar,
Tony Wu,
George Yiu,
Anthony D. Joseph, and
J. D. Tygar}

\authorrunning{Brad Miller et al.}

\institute{
Google Inc.
\email{bradmiller@google.com}
\and
UC Berkeley
\email{\{akant,riyazdf,vaishaal,tony.wu,adj,tygar\}@cs.berkeley.edu}
\and
International Computer Science Institute
\email{\{mct,sadia\}@icsi.berkeley.edu}
\and 
Netflix
\email{rbachwani@netflix.com}
\and
DataVisor
\email{ling.huang@datavisor.com}
\and
Pinterest
\email{george@pinterest.com}
}

\maketitle

\subsection*{Abstract}

\newcommand\blfootnote[1]{%
  \begingroup
  \renewcommand\thefootnote{}\footnote{#1}%
  \addtocounter{footnote}{-7}%
  \endgroup
}

We present and evaluate a large-scale malware detection system integrating machine learning with expert reviewers, treating reviewers as a limited labeling resource. 
We demonstrate that even in small numbers, reviewers can vastly improve the system's ability to keep pace with evolving threats. 
We conduct our evaluation on a sample of VirusTotal submissions spanning 2.5 years and containing 1.1 million binaries with 778GB of raw feature data.
Without reviewer assistance, we achieve 72\% detection at a 0.5\% false positive rate, performing comparable to the best vendors on VirusTotal. 
Given a budget of 80 accurate reviews daily, we improve detection to 89\% and are able to detect 42\% of malicious binaries undetected upon initial submission to VirusTotal. 
\blfootnote{\textsuperscript{\textdagger}Primary contribution while at UC Berkeley. \textsuperscript{\textdaggerdbl}Primary contribution while at Intel.}
Additionally, we identify a previously unnoticed temporal inconsistency in the labeling of training datasets.
We compare the impact of training labels obtained at the same time training data is first seen with training labels obtained months later.
We find that using training labels obtained well after samples appear, and thus unavailable in practice for current training data, inflates measured detection by almost 20 percentage points.
We release our cluster-based implementation, as well as a list of all hashes in our evaluation and 3\% of our entire dataset.

\section{Introduction}
\label{sec:intro}

Malware constitutes an enormous arms race in which attackers evolve to evade detection and detection mechanisms react.
A recent study found that only 66\% of malware was detected within 24 hours, 72\% within one week, and 93\% within one month~\cite{damballa}.
To evade detection, attackers produce a large number of different malware binaries, with McAfee receiving over 300,000 binaries daily~\cite{McAfeeQ2_2014}.

Machine learning offers hope for timely detection at scale, but the setting of malware detection differs from common applications of machine learning.
Unlike applications such as speech and text recognition where pronunciations and character shapes remain relatively constant over time, malware evolves as adversaries attempt to fool detectors.
In effect, malware detection becomes an online process in which vendors must continually update detectors in response to new threats, requiring accurate labels for new data.
Unfortunately, malware labeling poses unique challenges.
Whereas reading is sufficient to label text, the deceptive and technical nature of malware requires expert analysis.

We present an approach to detection integrating machine learning and expert reviews to keep pace with new threats at scale.
As expert labeling is expensive, we model the expert as capable of supplying labels for a limited selection of samples.
We then combine the limited supply of expert reviewer labels with the broader supply of noisy labels produced by anti-virus scanners to train a detection model.
We evaluate our approach using a sample of submissions to VirusTotal, a malware analysis and detection website~\cite{virustotal}.
The dataset includes a timestamp and anti-virus labels for each submission, capturing the emergence and prevalence of binaries, as well as label knowledge, over a 2.5 year period.
We train new models weekly with a customized approach combining accurate reviewer labels and noisy anti-virus labels and evaluate each model over the coming week.
To evaluate at scale, we simulate reviewer labels by revealing the results of automated scans taken at least 8 months after a sample first appears, providing opportunity for automated detectors to update and detect new threats.

We recognize that accurate training labels are not instantaneously available for all data, and therefore examine the impact of training label practices on performance measurement.
Prior work has introduced \textit{temporal sample consistency}, requiring that training binaries predate evaluation binaries~\cite{kolter06}.
We introduce \textit{temporal label consistency}, imposing the requirement that training labels also predate evaluation binaries.
Temporal label consistency restricts label quality relative to common practice, which collects labels well after binaries first appear and uses the same mature labels for both training and evaluation, leading to artificially inflated performance measurements.

Our work offers the following contributions:
\begin{itemize}

\item We present a detection system that integrates reviewers to increase detection from 72\% at 0.5\% false positive rate, comparable to the best vendors on VirusTotal, to 77\% and 89\% detection with a budget of 10 and 80 reviews daily on average.  Additionally, our system detects 42\% of malicious binaries initially undetected by vendors in our evaluation.

\item We demonstrate impact of temporally inconsistent labels on performance measurement, artificially inflating measured detection from 72\% to 91\% at a 0.5\% false positive rate.

\item We publicly release\footnote{\url{http://secml.cs.berkeley.edu/detection_platform/}} our implementation, 3\% of all data, and list of all 1.1 million unique binaries appearing over 2.5 years included in our evaluation.

\end{itemize}

Our evaluation also includes several additional experiments offering a more complete understanding of detection performance.
Although our design includes both static and dynamic features, since VirusTotal detectors must operate statically we also compare our performance against VirusTotal using static features alone.
Note that the restriction to static features actually disadvantages our approach, as VirusTotal detectors may operate against the arbitrary file and we restrict ourselves to static attributes available through VirusTotal.
Our performance is slightly impacted, producing 84\% detection at 0.5\% false positive rate with 80 queries daily and still surpassing detectors on VirusTotal.
We also explore the impact of inaccurate human labelers on the system's detection performance by adding random noise to the simulated expert labels.
We find that our design is robust in the presence of imperfect labelers.
Given reviewers with a 90\% true positive rate and a 5\% false positive rate our system still achieves 82\% detection at a 0.5\% false positive rate, as compared to 89\% detection using accurate reviewers.

We evaluate our contributions using VirusTotal data because each submission represents a request for analysis from a user, researcher or member of the security community.
VirusTotal responds to requests by running dozens of anti-virus products from the security industry, including large firms such as McAfee, Symantec and Kaspersky.
As we evaluate our contributions on a dataset including submissions from researchers and the security industry, not a random sampling of files from end user machines, we envision our approach as improving the detection workflows within security firms which ultimately produce products for end users.
We demonstrate that by investing a fraction of the engineering expertise of large security firms, we can vastly improve the ability to determine whether a binary is malicious.

In Section~\ref{sec:related}, we review prior work.
Section~\ref{sec:design} presents the design of our system, including feature extraction, machine learning and integration of the labeling expert, and Section~\ref{sec:data} examines our dataset.
Section~\ref{sec:evaluation} discusses our system implementation and then examines the impact of different performance measurement techniques and evaluates the performance of our detection system.
Lastly, Section~\ref{sec:conclude} concludes.

\section{Prior Work}
\label{sec:related}

In this section we present the prior work most directly related to our own areas of contribution: reviewer integration to improve automated detection and performance measurement.
Consistent with the focus of our work, we primarily discuss systems for malware detection rather than family classification or clustering.
An extensive discussion of related work is available online~\cite{miller2015}.

Since minimal prior work has explored reviewer integration, we begin by discussing systems that moderate access to any expensive labeling resource.
Several works employ a \textit{weak detector} design, which cheaply labels \textit{some} instances as benign but requires an expensive confirmation to label \textit{any} instance as malicious.
Provos~et~al. and Canali~et~al. present weak detector systems for malicious URLs which moderate access to expensive analysis in a virtual machine~\cite{provos2008, canali2011}.
Similarly, Chakradeo~et~al. present MAST, a system capable of detecting 95\% of Android malware at the cost of analyzing 13\% of non-malicious applications~\cite{chakradeo2013}.
Karanth~et~al. prioritize JavaScript for manual review with the end goal of identifying new vulnerabilities~\cite{karanth2011}.
In contrast with weak detectors, we view the expensive resource as an integrated component in a periodically retrained system, rather than the final step in a detection pipeline.
Instead of attempting to pass the entire and exact set of malicious instances to the expensive resource for verification, we identify a smaller set of instances that improve automated detection and use scalable components to determine final instance labels.

In contrast to weak detector approaches, Nissim~et~al. present a system that integrates reviewers during retraining but focuses on increasing the raw number of malicious instances submitted to the reviewer rather than improving automated detection.
Nissim~et~al. introduce two reviewer integration strategies and compare both to \textit{uncertainty sampling}, a reviewer integration technique from machine learning~\cite{settles2009}.
Although each new strategy reviews more malicious samples, neither improves automated detection, instead producing lower true positive and higher false positive rates~\cite{nissim2014pdf} or true positive rates within 1\%~\cite{nissim2014al} of uncertainty sampling.
The evaluation also lacks timestamped data and randomly divides samples into 10 artificial ``days''.
Since there are no temporal effects in the sample ordering, it is not possible to accurately assess detector performance or reviewer workload when confronted with new attacks.
In contrast, we demonstrate novel reviewer integration improving detection 17 percentage points over uncertainty sampling and conduct an evaluation with timestamped samples and labels spanning 2.5 years.

Sculley~et~al. present Google's approach to detecting adversarial advertisements, integrating human reviewers and automated detection~\cite{sculley2011}.
Unfortunately, the presentation omits key details and the sensitive nature of the system prevents any code or data release.
For example, the evaluation does not specify how many human reviewers are necessary, the added benefit from additional reviewers or the total number of queries to each reviewer.
Likewise, the impact of reviewers errors and different integration strategies is also unspecified.
We contribute an analysis of the marginal benefit from additional reviews, as well as the impacts of reviewer errors and different reviewer integration strategies.
Additionally, we release all source code and sample data to facilitate future work.

We also examine prior work related to performance measurement.
The most common performance measurement technique in malware detection is \textit{cross-validation}~(e.g.,~\cite{curtsinger2011, schultz2001dmm, stringhini2013, arp2014}).
Cross-validation tends to inflate measured performance by partitioning training and evaluation data randomly, effectively guaranteeing that any attack seen in evaluation is also seen in training~\cite{kantchelian2013}.
Kolter~et~al. improve on cross-validation by using a separate training dataset which entirely predates any evaluation data~\cite{kolter06}.
Furthering this approach, Perdisci~et~al. and Srndic~et~al. conduct evaluations which use a single timestamped dataset divided chronologically into periods, using the first $n-1$ periods to detect content in period $n$~\cite{perdisci2010, srndic2013}.
While these works maintain temporal sample consistency, none present or systematically evaluate the impact of temporal label consistency.

Prior work approaching temporal label consistency has either evaluated a system in production, which would have no way to be temporally inconsistent, or a system that retrains on its own output.
Rajab~et~al. evaluate a deployed PDF malware detector, which trains using presently available knowledge and is evaluated in retrospect after anti-virus labels have matured~\cite{rajab2013}.
Schwenk~et~al. demonstrate the infeasibility of a JavaScript malware system which is iteratively retrained over time using its own output labels, but do not compare temporally consistent labels from an external source with labels from the future~\cite{schwenk2012}.

\begin{figure*}[t]
\begin{center}
\includegraphics[width=\textwidth]{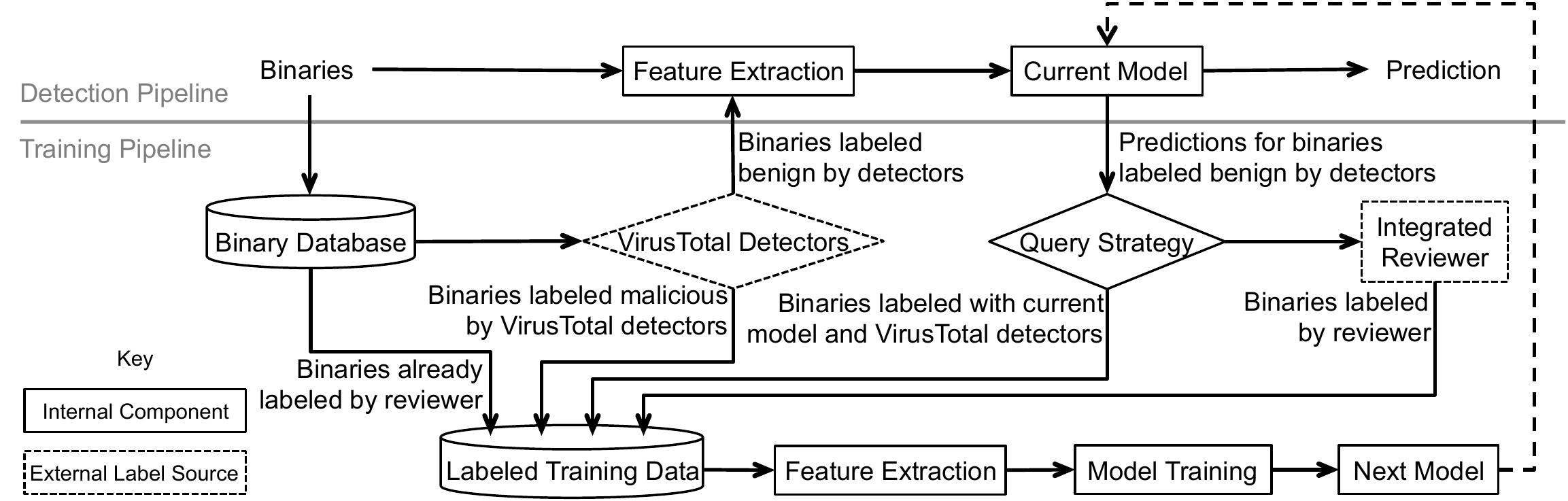}
\end{center}
\caption{The detection pipeline employs the current model to detect malware, and the training pipeline produces the next model for use in the detection pipeline.  During each retraining period, the training pipeline reviews all available training data and selects binaries for submission to the integrated reviewer.  Binaries labeled by the reviewer are combined with binaries labeled using the current model and anti-virus scan results to train the next model.}
\label{fig:design}
\end{figure*}

\section{Detector Design}
\label{sec:design}
In this section we present our detector design, including feature extraction, machine learning and reviewer integration.
Figure~\ref{fig:design} presents an overview of our approach.
When a binary arrives, the detection pipeline extracts the features, applies the current model to classify the binary as malicious or benign, and the training pipeline stores the binary in a database along with all other binaries seen to-date.
During each retraining period, binaries not detected by scanners on VirusTotal are considered for submission to the integrated reviewer.
Binaries confidently detected by the current model are included in training data with a malicious label, and the remaining purportedly benign binaries are submitted to the integrated reviewer as the review budget allows.
The remaining un-submitted binaries are included in the training data as benign.
At the end of the retraining period, the next model produced in the training pipeline replaces the current model and the process repeats.

We begin by examining the general techniques used for feature vectorization in Section~\ref{subs:fext}, and then present the application of feature vectorization techniques to static and dynamic attributes of binaries in Sections~\ref{subs:staticdynamic}.
Section~\ref{sec:training-labels} presents our approach to labeling training data, and Section~\ref{sec:active} describes our approach to reviewer integration.

\subsection{Approaches to Feature Vectorization}
\label{subs:fext}
Many machine learning algorithms work best with numeric features, but not all attributes of binaries come in that format.
We discuss four general techniques to convert static and dynamic attributes of binaries into numerical feature vectors.
Which of the four techniques we can apply varies across attributes.
For each technique, we discuss how we apply the technique to maximize robustness against evasion.

\minisec{Categorical}
The categorical mapping associates one dimension with each possible attribute value.
For example, the \texttt{DeviceIoControl} API call may correspond to index $i$ in feature vector $x$, where $x_i = 1$ if and only if the binary issues the \texttt{DeviceIOControl} API call.
Since the absence of an attribute reveals information about a binary, we include a special \emph{null index} to indicate that the value of the attribute is missing.  
For example, the file may not generate any network traffic, or may not be signed.
Where possible, we structure our application of categorical feature extraction to constrain the attacker to remain within a limited set of values.
For example, we apply subnet masks to IP addresses accessed by binaries to effectively shrink the IP space and associate access to similar IP addresses with the same feature index.

\minisec{Ordinal}
Ordinal attributes assume a specific value in an ordered range of possibilities, such as the size of a binary.
To remain robust to moderate fluctuations as adversaries attempt to evade detection, we vectorize ordinal values using a binning scheme rather than associating each distinct quantity with a unique index.
The binning scheme works as follows: for a given attribute value, we return the index of the bin which the value falls into, and set the corresponding dimension to 1. 
For attributes that vary widely, we use a non-linear scheme to prevent large values from overwhelming small values during training.
For example, the number of written files $v$ is discretized to a value $i$ such that $3^i \leq v < 3^{i+1}$, where the exponential bins accommodate the large dynamic range of this quantity.  

\minisec{Free-form String}
Many important attributes appear as unbounded strings, such as the comments field of the signature check.
Representing these attributes as categorical features could allow an attacker to evade detection by altering a single character in the attribute, causing the attribute to map into a different dimension.
To increase robustness, we capture 3-grams of these strings, where each contiguous sequence of 3 characters represents a distinct 3-gram, and consider each of the 3-grams as a separate dimension. 
Since this approach is still sensitive to variations that alter 3-grams, we introduce an additional string simplification.

To reduce sensitivity to 3-gram variations, we define classes of equivalence between characters and replace each character by its canonical representative. 
For instance, the string \texttt{3PUe5f} would be canonicalized to \texttt{0BAa0b}, where upper and lowercase vowels are mapped to `\verb+A+' and `\verb+a+'  respectively, upper and lowercase consonants are mapped to `\verb+B+' and `\verb+b+', and numerical characters to `\verb+0+'. 
Likewise, the string \texttt{7SEi2d} would also canonicalize to \texttt{0BAa0b}.
Occasionally, we sort the characters of the trigrams to further control for variation and better capture the morphology of the string. 
Mapping portable executable resource names, which sometimes exhibit long random-looking bytes sequences, is one application of this string simplification technique.

\minisec{Sequential}
The value of some attributes is a sequence of tokens where each token assumes a finite range of values.
These sequential attributes are strongly related to free-form string attributes, although the individual tokens are not restricted to being individual characters.
We use sequential feature extraction to capture API call information since there is a finite set of API calls and the calls occur in a specific order.
As with free-form string features, we use an $n$-gram approach where each sequence of $n$ adjacent tokens comprises an individual feature.
Sequential vectorization can be vulnerable to evasion in situations where adversaries are able to introduce tokens which have no effect and separate meaningful tokens.
To increase robustness, we apply $n$-gram vectorization with $n=1$ and $n=2$ as well as $n=3$, decreasing the number of unique $n$-grams which the adversary is able to generate.

\begin{table*}[t] 
\resizebox{\linewidth}{!}{
\begin{tab}{1.05}{l@{\hspace{0.4em}}l l l}
&  \textbf{Feature Name} & \textbf{Description} & \textbf{Example} \\
\midrule
\multirow{6}{*}{\rotatebox{90}{\textbf{Static}}} & Binary Metadata* & Metadata from \textsc{MAGIC} and \textsc{EXIFTOOL} & \texttt{PECompact2 compressed}  \\
& Digital Signing* & Certificate chain identity attributes & \texttt{Google Inc}; \texttt{Somoto Ltd} \\ 
& Heuristic Tools & \textsc{trid}; Tools from ClamAV, Symantec & \texttt{InstallShield setup}; \texttt{DirectShow filter}  \\ 
& Packer Detection & Packer or crypter used on binary & \texttt{UPX}; \texttt{NSIS}; \texttt{Armadillo} \\ 
& PE Properties*\textsuperscript{\textdagger} & Section hashes, entropies;  Resource list, types & \texttt{image/x-png}; \texttt{hash:eb0c7c289436...} \\ 
& Static Imports &  Referenced library names and functions & \texttt{msvcrt.dll/ldiv}; \texttt{certcli.dll} \\ 
\midrule
\multirow{7}{*}{\rotatebox{90}{\textbf{Dynamic}}} & Dynamic Imports & Dynamically loaded libraries & \texttt{shell32.dll}; \texttt{dnsapi.dll} \\ 
& File Operations\textsuperscript{\textdagger} & Number of operations; File paths accessed & \texttt{C:\textbackslash{WINDOWS}\textbackslash{system32}\textbackslash{mshtml.tlb}} \\ 
& Mutex Operations* & Each created or opened mutex & \texttt{ShimCacheMutex}; \texttt{RasPbFile} \\ 
& Network Operations\textsuperscript{\textdagger} & IPs accessed; HTTP requests; DNS requests & \texttt{66.150.14.*}; \texttt{b.liteflames.com} \\ 
& Processes & Created, injected or terminated process names & \texttt{python.exe}; \texttt{cmd.exe} \\ 
& Registry Operations &  Registry key set or delete operations & \texttt{SET: ...\textbackslash{WindowsUpdate}\textbackslash{AU}\textbackslash{NoAutoUpdate}} \\ 
& Windows API Calls\textsuperscript{\textdaggerdbl}  & $n$-grams of Windows API calls & \texttt{DeviceIoControl | IsDebuggerPresent} \\ 
\end{tab}
}
\vspace{.4cm}
\caption{Feature vectors reflect static and dynamic attributes of binaries.  We apply categorical vectorization to all attributes, as well as *string, \textsuperscript{\textdagger}ordinal and \textsuperscript{\textdaggerdbl}sequential vectorization for selected attributes.}
\label{tab:feature}
\end{table*}

\subsection{Attributes of Binaries}
\label{subs:staticdynamic}
VirusTotal provides static and dynamic attributes for each binary.
Whereas static attributes are obtained though analysis of the binary itself, dynamic attributes are obtained through execution in the Cuckoo sandbox~\cite{cuckoosandbox}.
Table~\ref{tab:feature} provides an overview of static attributes, dynamic attributes and associated vectorization techniques.

The static attributes available from VirusTotal consist of direct properties of the executable code itself, metadata associated with or derived from the executable and the results of heuristic tools applied to the executable.
The attributes extracted directly from the code include any statically imported library functions and aspects of the portable executable format, such as resource language, section attributes (e.g. entropy) and resource attributes (e.g. type).
The metadata associated with the code includes the output of the \textsc{magic} and \textsc{exiftool} utilities, which infer properties such as the file type, and any digital signatures associated with the file.
We collect the status of the verification, the identities of every entity in the certificate chain, comments, product name, description, copyright, internal name, and publisher from each digital signature.
The heuristic tools applied to the executable include \textsc{peid}~\cite{peid} and utilities from ClamAV~\cite{clamavpua}, and check for packing, network utilities or administrative utilities commonly associated with malware or potentially unwanted applications.

The dynamic attributes available from the Cuckoo sandbox capture interactions with the host operating system, disk and network resources.
Interactions with the operating system include dynamic library imports, mutex activity and manipulation of other processes running on the system.
Additionally, the Cuckoo sandbox provides an execution trace of all Windows API calls accessed by the binary, including the arguments, argument values and return values of any system call.
The summary of disk activity includes file system and registry operations, capturing any persistent effects of the binary.
We utilize both full and partial paths of file system operations as well as the types and number of operations to the file system during feature extraction; we also utilize the specific registry keys accessed or modified by the binary.
Lastly, we extract features from the network activity of the binary, including HTTP and DNS traffic and IP addresses accessed via TCP and UDP.

\subsection{Training Label Harmonization and Reviewer Query Strategy}
\label{sec:training-labels}

During each retraining period, the training process must assign labels to all available training binaries.
The process of assigning training labels harmonizes four distinct sources of information: scan results from anti-virus software, the current learned model, any prior reviews, and additional fresh reviews for a small number of binaries selected by the \textit{query strategy} for review.

The labeling process begins with the anti-virus scan results and application of the current model, both of which prune the set of binaries which the query strategy will consider for submission to the integrated reviewer.
Our application of anti-virus scan results leverages the intuition, which we confirm in Section~\ref{sec:data}, that anti-virus vendors bias detections towards false negatives rather than false positives.
Correspondingly, we view consensus among anti-virus detectors that a binary is malicious as sufficient to label the binary malicious during training, but we do not label undetected binaries as benign without further analysis.
We call this heuristic the \textit{undetected} filter since only binaries which are not detected by the vendors remain as candidates for review.

Next, we apply our current detection model to all undetected binaries and assign a malicious label to any binaries which score above a threshold $M$.
We refer to this heuristic as \textit{auto-relabeling} since some undetected binaries are automatically relabeled, similar to the self-training concept from semi-supervised learning~\cite{chapelle2010}.
If the binary is both undetected by anti-virus vendors and cannot be auto-relabeled using our detector, we submit the binary to the query strategy.

From the binaries that could not be confidently labeled as malicious, the query strategy selects a subset for review to improve their training labels.
The \textit{uncertainty sampling} query strategy selects binaries that are closest to the decision boundary, intuiting that the model will benefit from knowing the labels of those binaries about which it is unsure~\cite{settles2009}.
Uncertainty sampling has experienced success in other application domains, such as text classification, and served as a baseline for comparison in prior work involving integrated manual review~\cite{nissim2014pdf, nissim2014al}.
Designed for a case where the reviewer is the only source of labeling information, uncertainty sampling is unaware of how our two heuristics used the noisy labels from anti-virus scanners to filter the binaries for its consideration.

Consequently, we propose a new query strategy aware of our heuristics to increase the effectiveness of the integrated reviewer.
Since the heuristics identify binaries likely to be malicious, we will label any binary not identified by them or selected for review as benign. 
Consequently, only reviews which label a binary malicious will impact the final training data labels.
Accordingly, we develop the \textit{maliciousness} query strategy, which selects binaries for review that received high scores from our detection model, but not high enough to be subject to auto-relabeling.
More formally, the query strategy has a submission budget $B$, where $B$ is determined as a fixed percentage of the total number of new training binaries during the retraining period.
The maliciousness query strategy then submits the $B$ remaining binaries with the greatest maliciousness scores less than the auto-relabeling threshold $M$ to the integrated reviewer.
The binaries in excess of $B$ which are not submitted to the integrated reviewer are labeled benign.
By selecting binaries likely to be malicious but would otherwise be labeled benign, maliciousness achieves a higher likelihood than uncertainty sampling that the review will effect a change in training labels.

\subsection{Model Training and Integration of Reviewer Labels}
\label{sec:active}

After considering several forms of learning, including decision tree and nearest neighbor based approaches, we selected logistic regression as the basis for our malware detector.
As a linear classifier, logistic regression assigns a weight to each feature and issues predictions as a linear function of the feature vector, resulting in a real valued quantity~\cite{hastie01}.
Scoring each binary as a real valued quantity enables us to create a tradeoff between true and false positive rates by adjusting the threshold at which binaries are labeled malicious.
Linear classification scales well in prediction as the size of the model is a function of the dimensionality of the data and not the size of the training data, as happens with nearest neighbor techniques.
Additionally, the clear relationship between weights and features allows analysts to easily understand what the detector is doing and why, which can be difficult with complex tree ensembles.
Lastly, logistic regression scales well in training with many available implementations capable of accommodating high dimensional feature spaces and large amounts of training data.

We now discuss our training process integrating labels from the reviewer with noisy labels from anti-virus scanners and our own detector.
Since the reviewer only labels a small minority of binaries, noisy labels from anti-virus vendors will overwhelm reviewer labels during training unless reviewer labels receive special treatment.
We present the standard logistic regression training process below, and then describe the special treatment which we provide for reviewer labels.
The logistic regression training process finds the weight vector $\mathbf{w}$ which minimizes the following loss function for labeled training set $\left\{(\mathbf{x}^1, y^1), \dots, (\mathbf{x}^n, y^n)\right\}$ where $y^i \in \{-1, +1\}$ represents the label:
 \newcommand{\ps}{\quad\!}
\[
C_- * \sum_{i:y^i=-1} \ell(-\mathbf{w}^\intercal \mathbf{x}^i) 
 \ps+\ps
C_+ * \sum_{i:y^i=1} \ell(\mathbf{w}^\intercal \mathbf{x}^i)
 \ps+\ps 
\frac{1}{2}\|\mathbf{w}\|^2
\]
$C_- > 0$ and $C_+ > 0$ are distinct hyper-parameters controlling for both regularization and class importance weighting and $\ell(x) = \log(1+\exp(-x))$ is the logistic loss function.
The first and second terms correspond to the misclassification losses for negative and positive instances, respectively, and the final term is a regularization term that discourages models with many large non-zero weights.
To amplify the effect of reviewer labels, we assign a higher weight $W$ during training to any binary labeled benign by the reviewer.
We obtain superior results only weighting binaries that the reviewer labels benign since the maliciousness query strategy tends to select binaries for review which fall on the malicious side of the decision boundary.
When a benign instance is classified as malicious during training, a particularly high weight is necessary to have a corrective effect on the model and force the instance to receive a benign classification.
\section{Dataset and Evaluation Labeling Overview}
\label{sec:data}

We maintain that an evaluation dataset should include diverse binaries, reflect the emergence and prevalence of binaries over time, and record changes in the best available labeling knowledge for the binaries as time progresses.
Our evaluation dataset, consisting of 1.1 million distinct binaries submitted to VirusTotal between January 2012 and June 2014, achieves these criteria.
VirusTotal accepts submissions from end users, researchers and corporations, leading to a diverse sampling of binaries containing thousands of malware families and benign instances.
To randomize interaction with daily and hourly batch submission jobs, VirusTotal supplied us with the hashes of binaries submitted during a randomized segment during each hour of our collection period, reflecting approximately 1\% of the total binaries during the collection period.
We include each submission of each binary to accurately represent the prevalence and labeling knowledge of binaries over time.
A more complete discussion of the dataset, including changes in vendor labels over time and analysis of our labeling methodology is available online~\cite{miller2015}.

Due to the regular distribution of the evaluation data over an extended period of time and the broad use of VirusTotal, the dataset includes a diverse sampling from many families of malware.
Symantec, TrendMicro, Kaspersky and McAfee report 3{,}135, 46{,}374, 112{,}114 and 408{,}646 unique families for the dataset, respectively.
The number of families reported varies due to differences in naming conventions between vendors.
Although the exact number of families reported varies by vendor, each vendor agrees that the malware represents a broad sampling, with each vendor reporting less than 50\% of malware occurring in the most common 10\% of families.

As the dataset contains scan results form 80 different vendors, we employ a harmonization approach to create the \textit{gold labels} which we use to characterize the dataset and evaluate detector performance.
Since some vendors are only sporadically present in the data, we restrict our work to the 32 vendors present in at least 97\% of scan results to increase consistency in the set of vendors applied to each binary.\footnote{In particular, we include the following vendors: AVG, Antiy-AVL, Avast, BitDefender, CAT-QuickHeal, ClamAV, Comodo, ESET-NOD32, Emsisoft, F-Prot, Fortinet, GData, Ikarus, Jiangmin, K7AntiVirus, Kaspersky, McAfee, McAfee-GW-Edition, Microsoft, Norman, Panda, SUPERAntiSpyware, Sophos, Symantec, TheHacker, TotalDefense, TrendMicro, TrendMicro-HouseCall, VBA32, VIPRE, ViRobot and nProtect.}
We observe that among binaries that receive multiple scans in our dataset, 29.6\% of binaries increase in number of detections as malware by at least 5 vendors from their first to last scan, and only 0.25\% of binaries decrease by 5 or more detections.
This shift from benign to malicious labels confirms the intuition that vendors behave conservatively, preferring false negatives over false positives.
Given vendors' demonstrated aversion to false positives, we set a detection threshold of 4 vendor detections as sufficient to label a binary as malicious, and request a rescan of any binary which received fewer than 10 detections at the most recent scan.
We conduct rescans in February and March 2015, 8 months after the end of our data collection period, to allow time for vendor signature updates.
We avoid rescanning binaries with 10 or more detections since decreases large enough to cross the four vendor detection threshold are unlikely.
After rescanning, we assign a gold label to each binary in our dataset representing the best available understanding of whether the binary is malicious.

\begin{figure*}[t]
	\centering
	\begin{subfigure}[]{.49\textwidth}
	\includegraphics[width=\textwidth]{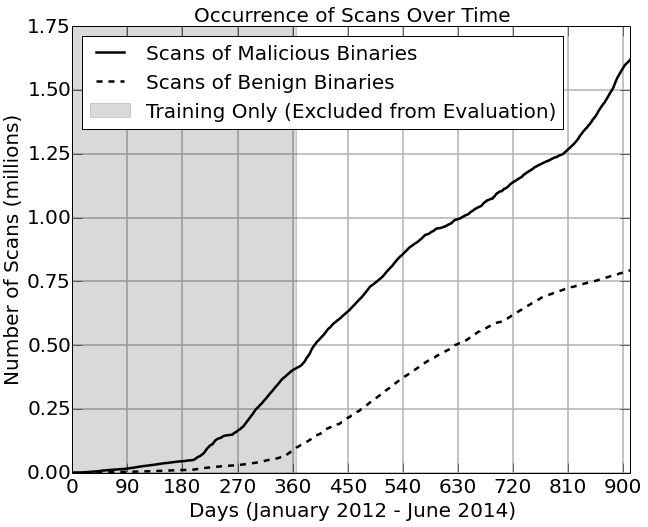}
	\caption{}
	\label{fig:time}
	\end{subfigure}
        \hfill
	\begin{subfigure}[]{.49\textwidth}
	\includegraphics[width=\textwidth]{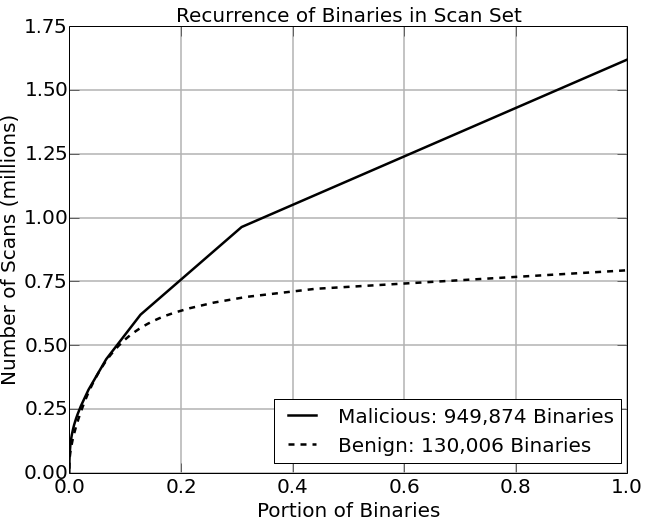}
	\caption{}
	\label{fig:repeats}
	\end{subfigure}
	\caption{Data Overview.  Figures~\ref{fig:time} and~\ref{fig:repeats} demonstrate that scans are well distributed across our evaluation period and distinct binaries, respectively.  Note that relative scarcity of scans in the first 200 days reflects availability of necessary attributes in VirusTotal data, not underlying submission behavior.  
	}
\end{figure*}

We reserve from January 2012 to December 2012, the first year of our data set, for obtaining an initial model and use the data from January 2013 to June 2014 to perform a complete rolling window evaluation of our detector.
Figure~\ref{fig:time} presents the occurrence of scans over time, indicating that scans consistently occur throughout the period during which we measure performance.
Notice that scans do not occur evenly during the training period, with the first approximately 200 days containing fewer scans.
The difference in available data occurs because fewer binaries have dynamic attributes available; the difference does not reflect an underlying phenomenon in submissions.

In addition to being well distributed over time, scans are also well distributed across the different binaries in our dataset.
Figure~\ref{fig:repeats} depicts the impact of resubmissions on the dataset, with the horizontal axis ordering binaries from most commonly to least commonly submitted.
We include re-submissions to ensure that the distribution of our evaluation data mirrors the distribution of actual data submitted to VirusTotal by incorporating the prevalence of each individual file, effectively balancing any effects of polymorphism in the dataset.
Additionally, inclusion of rescan events in our analysis provides more timely labeling during evaluation.

\section{Experimental Results and System Evaluation}
\label{sec:evaluation}

In this section we briefly discuss our implementation, present experimental results and evaluate our detection system.
Our presentation of experimental results demonstrates the impact of different performance measurement techniques on detection results.
Our detection system evaluation demonstrates the potential for integrated review techniques to improve performance over current anti-virus vendors, as well as the impact of reviewer errors, marginal benefit of additional reviews and effects of different of reviewer integration strategies.

\subsection{System Implementation}

Since anti-virus vendors can receive in excess of 300,000 binaries daily~\cite{McAfeeQ2_2014}, we design our detector with a focus on scalability. 
We implement our detection platform in five thousand lines of Python, which offers bindings for the numerical and infrastructure packages we require.  
We use Scikit Learn and Numpy for machine learning, and Apache Spark for distributed computation.  
Using a 40 core cluster with 600GB of RAM, we were able to conduct feature vectorization, learning and prediction on our 778GB dataset including 1.1 million unique binaries in 10 hours.

To allow experimentation at scale, we simulate an integrated reviewer rather than employing an actual labeling expert.
We model the analysis of the integrated reviewer by revealing the gold label associated with a binary.
For experiments that consider an imperfect reviewer, we assign the simulated reviewer a true positive rate and a false positive rate, allowing the likelihood of the reviewer supplying the correct label to depend on the gold label for the sample.
By conditioning the likelihood of a correct response on the gold label of a sample, we are able to more closely model the errors of an actual reviewer who may be highly likely to correctly identify a benign binary as benign, but less likely to correctly identify a malicious binary as malicious.
We leave the comparison of this model to actual reviewer performance as future work.

Lastly, we describe our management of the system parameters discussed in Section~\ref{sec:design}, including a reviewer submission budget $B$, auto-relabeling confidence threshold $M$ and learning parameters $C_-$, $C_+$ and $W$.
Section~\ref{sec:results} presents the effects of varying the submission budget $B$, with experiments conducted at 80 queries daily on average unless otherwise specified.
The remaining parameters are tuned to maximize detection at false positive rates between .01 and .001 on a set of binaries obtained from an industry partner and excluded from our evaluation.
We use the following values: $M = 1.25$, $C_- = 0.16$, $C_+ = .0048$ and $W = 10$.

\subsection{Impact of Performance Measurement Techniques}
\label{sec:method}

The primary motivation for measuring the performance of a detection system in a research or development setting is to understand how the system would perform in a production setting.
Accordingly, measurement techniques should seek to minimize the differences from production settings.
In practice, knowledge of both binaries and labels changes over time as new binaries appear and malware detectors respond appropriately with updated labels.
Performance measurement techniques that fail to recognize the emergence of binaries and label knowledge over time effectively utilize knowledge from the future, inflating the measured accuracy of the approach.
For example, consider malware that evades detection but can be easily detected once the first instance is identified.
Performance inflation occurs because inserting correctly labeled binaries into training data circumvents the difficult task of identifying the first instance of the malware.

\begin{figure*}[t]
	\centering
	\begin{subfigure}{.55\textwidth}
	\includegraphics[width=\linewidth]{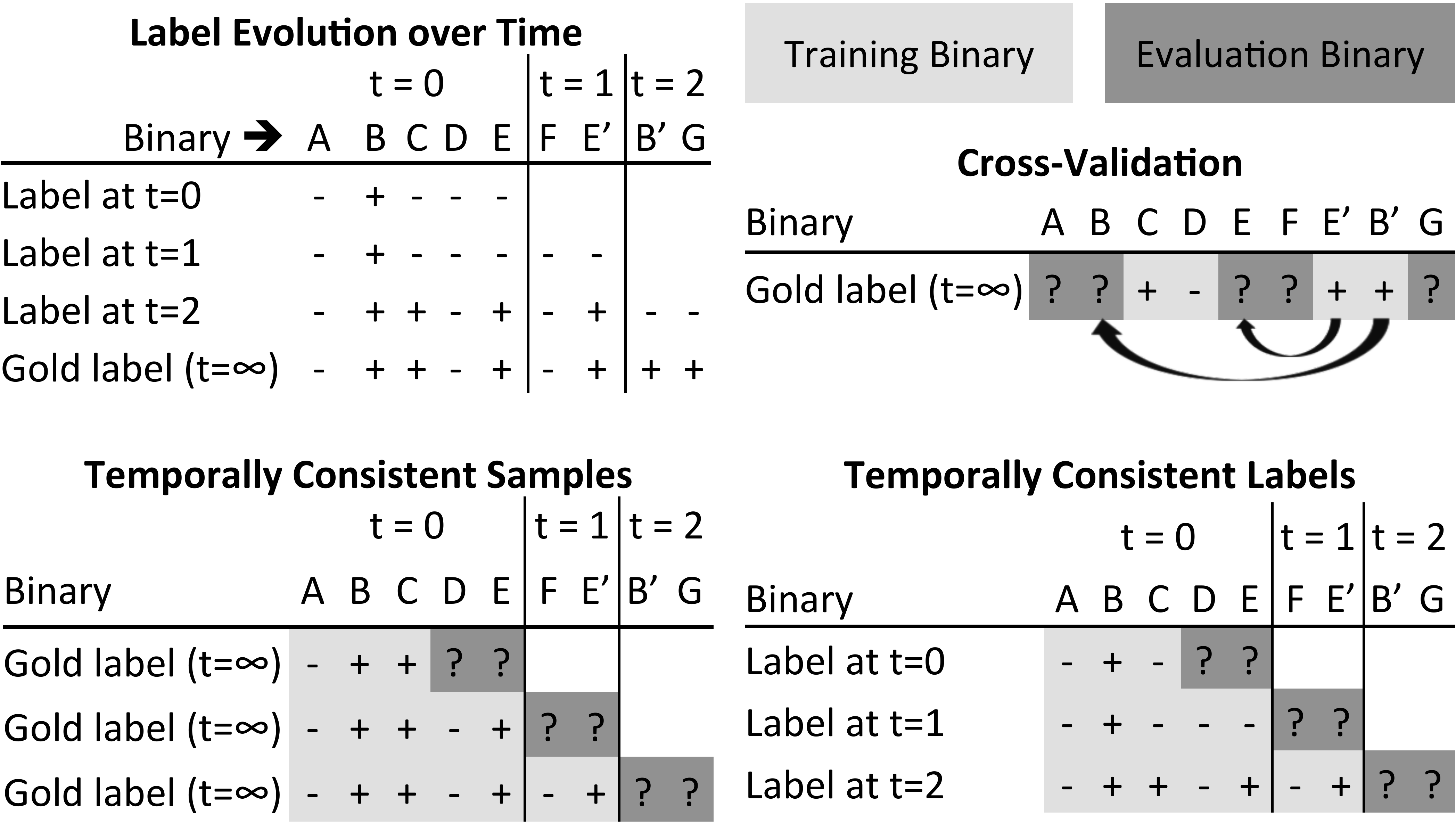}
	\caption{}
	\label{fig:style}
	\end{subfigure}
	\hfill
	\begin{subfigure}{.4\textwidth}
	\includegraphics[width=\linewidth]{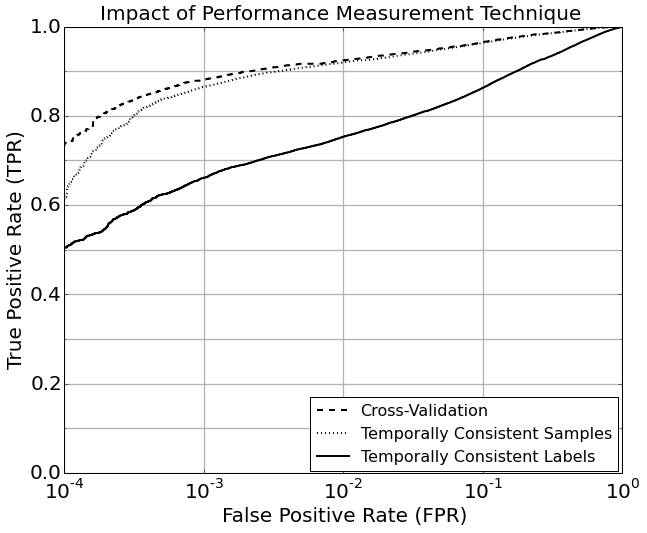}
	\caption{}
	\label{fig:xval}
	\end{subfigure}
	\caption{
Accurate performance measurement requires temporally consistent labels.
Figure~\ref{fig:style} illustrates three techniques.
The upper left shows the evolution of labels over time for a series of binaries, with B' and E' denoting variants of previously submitted binaries B and E.
Each remaining subfigure depicts the experiments a performance measurement technique would conduct given the example dataset.
Rows correspond to successive retraining periods with specified training and evaluation data, binaries appear chronologically from left to right, and \texttt{+} and \texttt{-} denote malicious and benign labels, respectively.
Figure~\ref{fig:xval} presents the effects of performance measurement technique on experimental results.}
	\label{fig:methodology}
\end{figure*}

We analyze three approaches to measuring the performance of malware detectors, each recognizing the emergence of binaries and labels over time to varying degrees.
Cross-validation is a common approach for machine learning evaluations in situations where binaries are independent and identically distributed (i.i.d.).
In the malware detection context the i.i.d.\ assumption does not hold since malware changes over time to evade detection.
Cross-validation evaluations completely disregard time, dividing binaries randomly and applying evaluation quality labels to all binaries.
Evaluations maintaining temporally consistent samples recognize the ordering of binaries in time but not the emergence of labels over time, instead applying gold labels from future scan results to all binaries.
Use of gold quality labels during training effectively assumes that accurate detection occurs instantly.
Evaluations maintaining temporally consistent labels fully respect the progression of knowledge, ordering binaries in time and restricting the training process to binaries and labels available at the time of training.
For measurements with both temporally consistent samples and labels, we divide data into periods and use the first $n-1$ periods to detect content in period $n$.
Unless otherwise specified we use a period length of one week.
Figure~\ref{fig:style} presents the specifics of each approach.

Our experiments demonstrate that measurement technique powerfully impacts performance results.
Figure~\ref{fig:xval} presents the results of our analysis.
Notice that cross-validation and temporally consistent samples perform similarly, inflating detection results 20 and 19 percentage points respectively over temporally consistent labeling at a 0.5\% false positive rate.
Since reviewer integration effectively reduces the impact of temporally consistent labels by revealing future labels, we conduct these experiments without any reviewer queries.
Note that our conclusions apply only to the setting of malware detection and not family classification, which presents a fundamentally different challenge as the set of known family labels may change over time.

Temporally consistent labeling requires that training labels predate evaluation binaries.
Since VirusTotal scans each binary upon each submission our experiments are able to satisfy temporally consistent labeling requirements.
However, since binaries are not necessarily rescanned at regular intervals, we are not able to guarantee that our labels are up to date.
For example, consider a binary which receives benign scan results in week 1 and malicious scan results in week 10: the up-to-date training label in week 5 is unclear.
To simulate the effects of more frequent rescanning, we conduct a second experiment in which we reveal the gold label for each binary once a fixed interval has passed since the binary's first submission.
We find that without releasing gold labels temporally consistent evaluation results in 76\% detection at a 1\% false positive rate; releasing gold labels 4 weeks and 1 week after a binary appears increases detection to 80\% and 84\% respectively.
Note that these figures represent an upper bound on the impact of frequent rescanning since malware may remain undetected much longer than 1 or 4 weeks.
Considering that cross-validation and temporal sample consistency each achieve 92\% detection at a 1\% false positive rate, we see that even with regular rescanning, temporal label consistency impacts detection results.

\subsection{Detection System Evaluation}
\label{sec:results}

In this section we evaluate our malware detection system and the impact of reviewer integration.
We begin with the impact of the reviewer and performance relative to VirusTotal.  Then, we examine parameters such as reviewer accuracy and retraining frequency.
Lastly, we analyze impact of different types of features.

\begin{figure}[t]
	\centering
	\includegraphics[width=.65\linewidth]{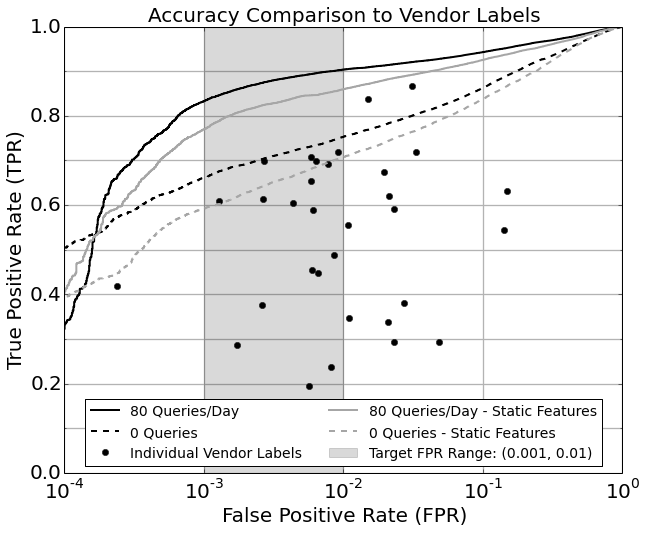}
	\caption{Without reviewer integration our detector is competitive with VirusTotal detectors.  With reviewer integration, detection improves beyond vendors on VirusTotal.  We tune our system to maximize detection in the (0.1\%, 1\%) false positive region, consequently decreasing detection at lower false positive rates.}
	\label{fig:compare_same_truth}
\end{figure}

\minisec{Impact of Integrated Reviewer}
Given the breadth of our data and unique structure of our evaluation, the vendor detection results on VirusTotal provide the best performance comparison for our work.
Based on the false positive rates of vendors, we tune our detector to maximize detection for false positive rates greater than 0.1\% and less than 1\%.
Figure~\ref{fig:compare_same_truth} compares our performance to vendor detectors provided on VirusTotal.
Without involvement from the integrated reviewer our detector achieves 72\% detection at a 0.5\% false positive rate, performing comparably to the best vendor detectors.
With support from the reviewer, we increase detection to 89\% at a 0.5\% false positive rate using 80 queries daily on average.
Since we train a separate model during each weekly retraining period, the performance curve results from varying the same detection threshold across the results of each individual model.

VirusTotal invokes vendor detectors from the command line rather than in an execution environment, allowing detectors to arbitrarily examine the file but preventing observation of dynamic behavior.
Since our analysis includes dynamic attributes, we also observe our performance when restricted to static attributes provided by VirusTotal.
Note that this restriction places our detector at a strict disadvantage to vendors, who may access the binary itself and apply signatures derived from dynamic analysis.
Figure~\ref{fig:compare_same_truth} demonstrates that our performance decreases when restricted to static features but, with support from the integrated reviewer, surpasses vendors to achieve 84\% detection at a 0.5\% false positive rate.

Performance comparison must also consider the process of deriving gold labels, which introduces a circularity that artificially inflates vendor performance.
Consider the case of a false positive: once a vendor has marked a binary as positive, the binary is more likely to receive a positive gold label, effectively decreasing the false positive rate of the vendor.
An alternate approach would be to withhold a vendor's labels when evaluating that vendor, effectively creating a separate ground truth for each vendor.
Although this approach more closely mirrors the evaluation of our own detector (which does not contribute to gold labels), in the interest of consistency we elect to use the same ground truth throughout the entire evaluation since efforts to correct any labeling bias only increase our performance differential.

In addition to offering superior detection performance aggregated across all data relative to vendor labels, our approach also experiences greater success detecting novel malware that is missed by detectors on VirusTotal.
Of the 1.1 million samples included in our analysis, there are 6,873 samples which have a malicious gold label but are undetected by all vendors the first time the sample appears.
Using 80 reviewer queries daily, our approach is able to detect 44\% and 32\% of these novel samples at 1\% and .1\% false positive rates, respectively.
The ability of our approach to detect novel malware illustrates the value of machine learning for detecting successively evolving generations of malware.

To provide a corresponding analysis of false positives, we measure our performance on the 61,213 samples which have a benign gold label and are not detected as malware by any vendor the first time the sample appears.
Of these 61,213 benign samples, our detector labels 2.0\% and 0.2\% as malicious when operating at 1\% and .1\% false positive rates over all data, respectively.
The increased false positive rate on initial scans of benign samples is expected since the sample has not yet been included as training data.

\begin{figure*}[t]
	\centering
	\begin{subfigure}{.45\textwidth}
	\includegraphics[width=\linewidth]{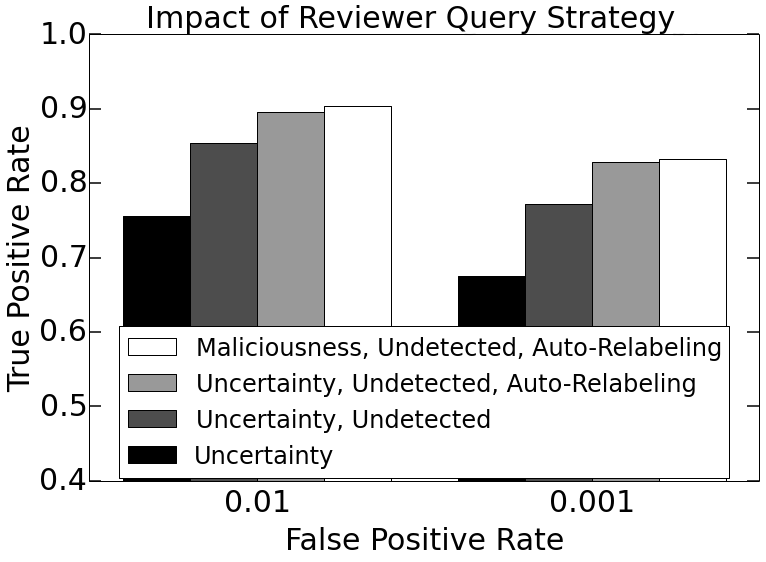}
	\caption{}
	\label{fig:strategies}
	\end{subfigure}
	\begin{subfigure}{.45\textwidth}
	\includegraphics[width=\linewidth]{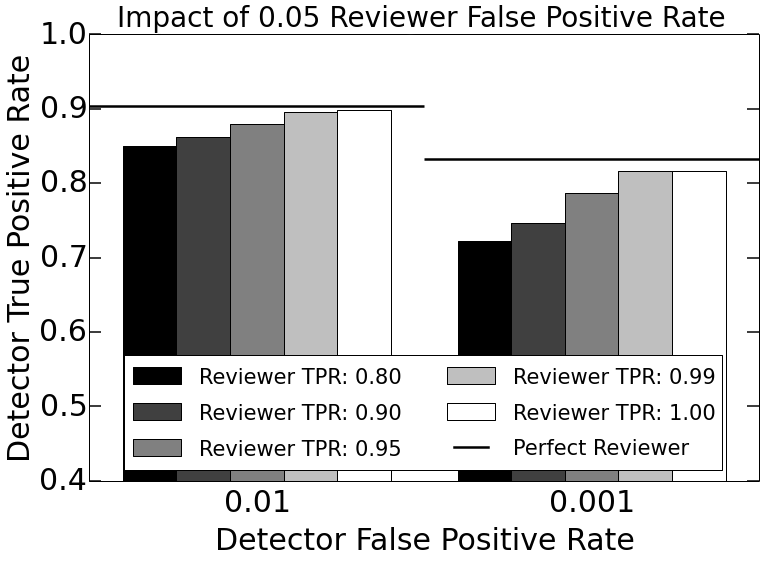}
	\caption{}
	\label{fig:error_05}
	\end{subfigure}
\caption{
Figure~\ref{fig:strategies} presents the impact of each component in our customized query strategy.  We improve detection over the uncertainty sampling approach from prior work. Figure~\ref{fig:error_05} presents the performance of our detector for imperfect reviewers with the specified true and false positive rates.  For example, given a reviewer with a 5\% false positive rate and 80\% true positive rate, our detector's true positive rate only decreases by 5\% at a 1\% false positive rate. 
}
\label{fig:oracle_design}
\end{figure*}

\minisec{Reviewer Query Strategies}

Our reviewer query strategy represents numerous advances over prior work. Figure~\ref{fig:strategies} presents the impact of each of the three improvements we introduce and discussed in Section~\ref{sec:training-labels}.
For a fixed labeling budget $B=80$, uncertainty sampling results in a detection rate 17 percentage points lower than the combination of our techniques at 0.1\% false positive rate.

\minisec{Reviewer Accuracy}

Our system also demonstrates strong results in the presence of an imperfect reviewer.
Since malware creators may explicitly design malware to appear benign but benign software is less likely to appear malicious, we model the false positive and true positive rates of reviewers separately, reflecting a reviewer who is more likely to mistake malware for benign software than benign software for malware.
Figure~\ref{fig:error_05} presents detection rates for reviewers with a 5\% false positive rates and a range of true positive rates.
For example, given a reviewer with a 5\% false positive rate and 80\% true positive rate, our detector's true positive rate only decreases by 5\% at a 1\% false positive rate.

\minisec{Resource Parameterization}

Beyond classifier parameters, detection performance is also influenced by operator resources including reviewer query budget and retraining frequency.
We explore each of these parameters below.

\begin{figure*}[t]
	\centering
	\begin{subfigure}{.45\textwidth}
	\includegraphics[width=\linewidth]{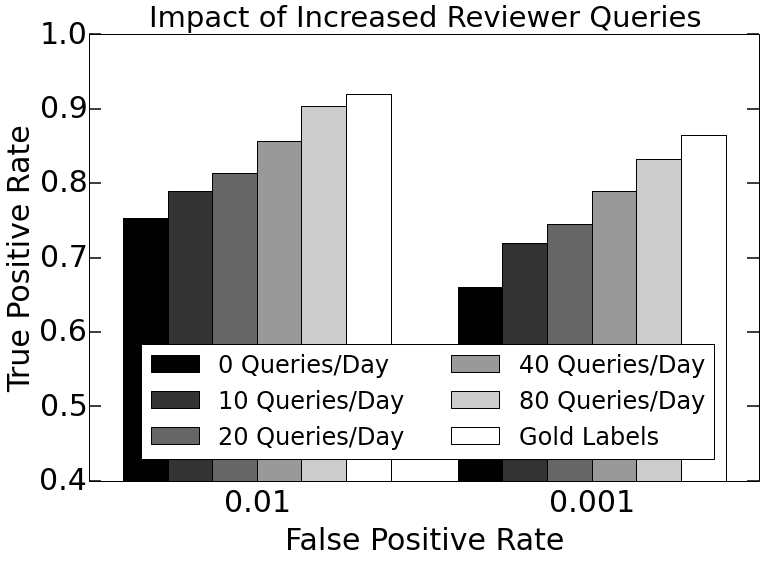}
	\caption{}
	\label{fig:oracle}
	\end{subfigure}
	\begin{subfigure}{.45\textwidth}
	\includegraphics[width=\linewidth]{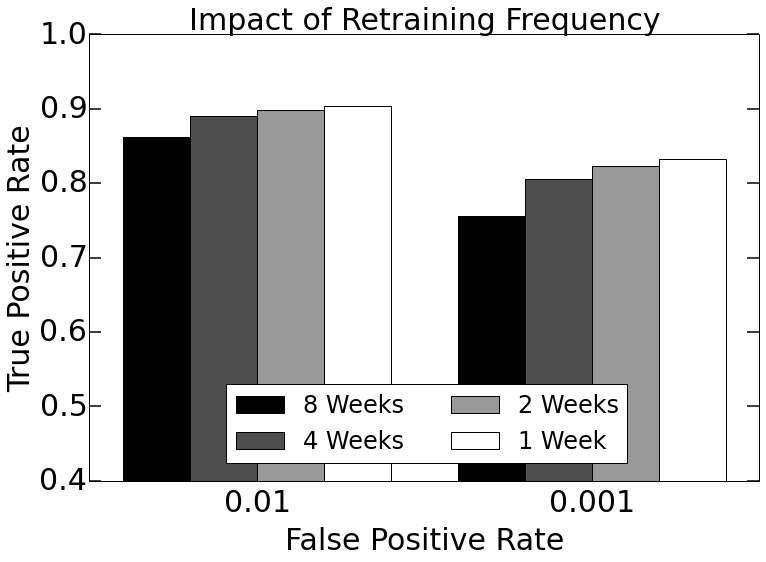}
	\caption{}
	\label{fig:lag}
	\end{subfigure}
\caption{
Figure~\ref{fig:oracle} presents performance for different reviewer query budgets, with significant return on minimal efforts and diminishing returns occurring around 80 queries/day.  
Figure~\ref{fig:lag} demonstrates that retraining more quickly improves detector performance.
}
\label{fig:oval}
\end{figure*}

As the allowed budget for queries to the reviewer increases, the detection performance increases since higher quality training labels are available.
Figure~\ref{fig:oracle} presents the detection increase from increased reviewer queries, with the benefit of 80 queries per day on average approaching the upper bound of having gold labels for all training data.
The benefit of reviewer queries is non-linear, with the initial queries providing the greatest benefit, allowing operators to experience disproportionate benefit from a limited review budget.

Although our evaluation is large relative to academic work, an actual deployment would offer an even larger pool of possible training data.
Since the utility of reviewer queries will vary with the size of the training data, increasing the amount of training data may increase reviewer queries required to reach full benefit.
Fortunately, the training process may elect to use only a subset of the available training data.
We demonstrate that 1.1 million binaries selected randomly from VirusTotal submissions is sufficient training data to outperform vendor labels for our evaluation data.

Lastly, we examine variations in the length of the re-training period governing how often models are updated.
We conduct these experiments with 80 reviewer queries on average per day.
Figure~\ref{fig:lag} presents the effect of variations in the retraining period.
Notice that the benefit of frequent retraining begins to diminish around 2 weeks.

\minisec{Detection Mechanics}

Having analyzed detection accuracy and evaluation methodology, we now examine the features that our detector uses for classification.
In the interest of understanding the dataset as a whole, we train a model over all data from all dates.
Although we learn a linear model and can easily observe the weight of each feature, inspecting the weight vector alone is not enough to understand feature importance.
A feature can be associated with a large weight but be essentially constant across the dataset, as may happen with strongly malicious features that are relatively rare in practice.
Intuitively, such features have low discrimination power. 
Furthermore, we are
interested in grouping low-level features together into high level
concepts.

Thus, we use the following ranking method for sets of features.
Let $d$ be the total number of features, $\mathbf{w} \in \mathbb{R}^d$ be the weight vector 
and $\{\mathbf{x}^i\}$ be a given set of instances. The notation $\mathbf{x}^i_k$ designates
the $k$-th coordinate of instance $\mathbf{x}^i$. We 
can compute the importance of a group $S \subset \{1,\dots,d\}$ of 
features by quantifying the amount of score variation $I_S$ they induce. 
Our ranking formula is:
\[
I_S = \sqrt{\Var_{i} \left[ \sum_{k \in S} 
\mathbf{x}^i_k \mathbf{w}_k \right]}
\]

\begin{figure}[t]
    \centering
    \includegraphics[width=.65\linewidth]{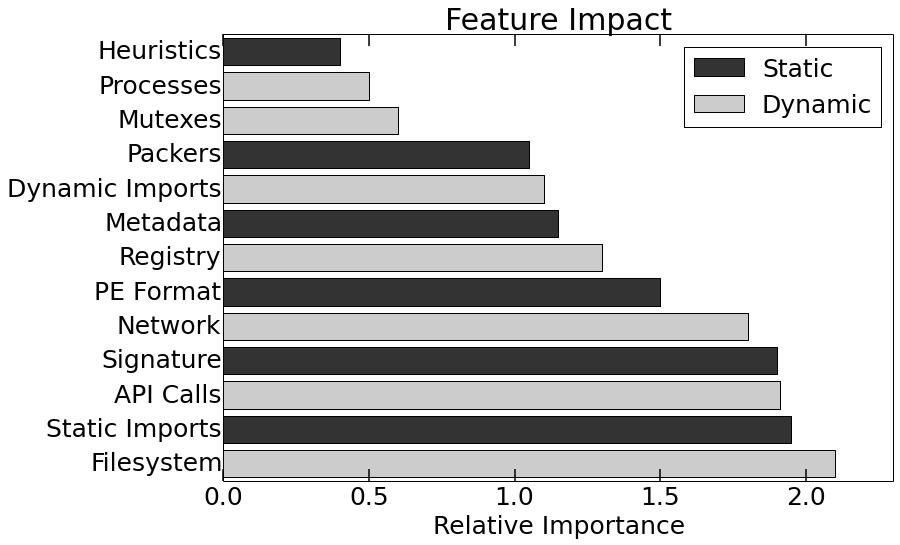}
    \caption{Feature categories ranked by importance.}
    \label{fig:ranked-categories}
\end{figure}

Using this ranking method,
Figure~\ref{fig:ranked-categories} shows the global ranking of the 
features when grouped by their original measurements. The most important
measurements are thus the file system operations, static imports, API call
sequence and digital signature, while the least useful measurement is the
heuristic tools.
Further analysis including highly weighted features is available online~\cite{miller2015}.

\section{Conclusion}
\label{sec:conclude}

In this paper, we explore the power of putting humans in the loop by integrating a simulated human labeling expert into a scalable malware detection system.
We show it capable of handling over 1 million samples using a small cluster in hours while substantially outperforming commercial anti-virus providers both in terms of malware detection and false positive rates (as measured using VirusTotal). 
We explain why machine learning systems appear to perform very well in research settings and yet fail to perform reasonably in production settings by demonstrating the critical temporal factors of labeling, training, and evaluation that affect detection performance in real-world settings. 
In future work, we plan to expand our detection system to perform malware family labeling and detection of new malware families.
Additionally, we may implement clustering or density based sampling techniques to further reduce the reviewer burden by eliminating any duplicate reviews.

\bibliographystyle{IEEEtranS}
\bibliography{pipeline}

\begin{thebibliography}{10}
\providecommand{\url}[1]{#1}
\csname url@samestyle\endcsname
\providecommand{\newblock}{\relax}
\providecommand{\bibinfo}[2]{#2}
\providecommand{\BIBentrySTDinterwordspacing}{\spaceskip=0pt\relax}
\providecommand{\BIBentryALTinterwordstretchfactor}{4}
\providecommand{\BIBentryALTinterwordspacing}{\spaceskip=\fontdimen2\font plus
\BIBentryALTinterwordstretchfactor\fontdimen3\font minus
  \fontdimen4\font\relax}
\providecommand{\BIBforeignlanguage}[2]{{%
\expandafter\ifx\csname l@#1\endcsname\relax
\typeout{** WARNING: IEEEtranS.bst: No hyphenation pattern has been}%
\typeout{** loaded for the language `#1'. Using the pattern for}%
\typeout{** the default language instead.}%
\else
\language=\csname l@#1\endcsname
\fi
#2}}
\providecommand{\BIBdecl}{\relax}
\BIBdecl

\bibitem{clamavpua}
``{ClamAV PUA},'' \url{http://www.clamav.net/doc/pua.html}, 11/14/14.

\bibitem{peid}
``{PEiD},'' \url{http://woodmann.com/BobSoft/Pages/Programs/PEiD}, 11/14/14.

\bibitem{cuckoosandbox}
``{The Cuckoo Sandbox},'' \url{http://www.cuckoosandbox.org}, 11/14/14.

\bibitem{arp2014}
D.~Arp, M.~Spreitzenbarth, M.~Hubner, H.~Gascon, and K.~Rieck, ``Drebin:
  Effective and explainable detection of android malware in your pocket.'' in
  \emph{NDSS}, 2014.

\bibitem{canali2011}
D.~Canali, M.~Cova, G.~Vigna, and C.~Kruegel, ``Prophiler: A fast filter for
  the large-scale detection of malicious web pages,'' in \emph{WWW 2011}.

\bibitem{chakradeo2013}
S.~Chakradeo, B.~Reaves, P.~Traynor, and W.~Enck, ``Mast: Triage for
  market-scale mobile malware analysis,'' in \emph{ACM WiSec}, 2013.

\bibitem{chapelle2010}
O.~Chapelle, B.~Schlkopf, and A.~Zien, \emph{Semi-Supervised Learning}.\hskip
  1em plus 0.5em minus 0.4em\relax The MIT Press, 2010.

\bibitem{curtsinger2011}
C.~Curtsinger, B.~Livshits, B.~Zorn, and C.~Seifert, ``Zozzle: Fast and precise
  in-browser javascript malware detection,'' in \emph{Usenix Security}, 2011.

\bibitem{damballa}
{Damballa}, ``{State of Infections Report: Q4 2014},'' Damballa, Tech. Rep.,
  2015.

\bibitem{hastie01}
T.~Hastie, R.~Tibshirani, and J.~Friedman, \emph{The Elements of Statistical
  Learning}.\hskip 1em plus 0.5em minus 0.4em\relax Springer New York Inc.,
  2001.

\bibitem{kantchelian2013}
A.~Kantchelian, S.~Afroz, L.~Huang, A.~C. Islam, B.~Miller, M.~C. Tschantz,
  R.~Greenstadt, A.~D. Joseph, and J.~D. Tygar, ``Approaches to adversarial
  drift,'' in \emph{ACM AISec}, 2013.

\bibitem{karanth2011}
S.~Karanth, S.~Laxman, P.~Naldurg, R.~Venkatesan, J.~Lambert, and J.~Shin,
  ``{ZDVUE:} prioritization of javascript attacks to discover new
  vulnerabilities,'' in \emph{ACM AISec}, 2011.

\bibitem{kolter06}
J.~Z. Kolter and M.~A. Maloof, ``Learning to detect and classify malicious
  executables in the wild,'' \emph{Journal of Machine Learning Research},
  vol.~7, 2006.

\bibitem{McAfeeQ2_2014}
{McAfee Labs}, ``{McAfee Labs Threats Report},'' August 2014.

\bibitem{miller2015}
B.~Miller, ``Scalable platform for malicious content detection integrating
  machine learning and manual review,'' Ph.D. dissertation, UC Berkeley, 2015.

\bibitem{nissim2014pdf}
N.~Nissim, A.~Cohen, R.~Moskovitch, A.~Shabtai, M.~Edry, O.~Bar-Ad, and
  Y.~Elovici, ``{ALPD}: Active learning framework for enhancing the detection
  of malicious pdf files,'' in \emph{IEEE JISIC}, Sept 2014.

\bibitem{nissim2014al}
N.~Nissim, R.~Moskovitch, L.~Rokach, and Y.~Elovici, ``Novel active learning
  methods for enhanced pc malware detection in windows os,'' in \emph{Journal
  of Expert Systems with Applications}, 2014.

\bibitem{perdisci2010}
R.~Perdisci, W.~Lee, and N.~Feamster, ``Behavioral clustering of http-based
  malware and signature generation using malicious network traces.'' in
  \emph{NSDI}, 2010.

\bibitem{provos2008}
N.~Provos, P.~Mavrommatis, M.~A. Rajab, and F.~Monrose, ``All your iframes
  point to us,'' in \emph{USENIX Security}, 2008.

\bibitem{rajab2013}
M.~A. Rajab, L.~Ballard, N.~Lutz, P.~Mavrommatis, and N.~Provos, ``{CAMP}:
  Content-{A}gnostic {M}alware {P}rotection.'' in \emph{NDSS}, 2013.

\bibitem{schultz2001dmm}
M.~G. Schultz, E.~Eskin, E.~Zadok, and S.~J. Stolfo, ``Data mining methods for
  detection of new malicious executables,'' in \emph{{IEEE S\&P}}, 2001.

\bibitem{schwenk2012}
G.~Schwenk, A.~Bikadorov, T.~Krueger, and K.~Rieck, ``Autonomous learning for
  detection of javascript attacks: Vision or reality?'' in \emph{ACM AISec},
  2012.

\bibitem{sculley2011}
D.~Sculley, M.~E. Otey, M.~Pohl, B.~Spitznagel, J.~Hainsworth, and Y.~Zhou,
  ``Detecting adversarial advertisements in the wild,'' in \emph{KDD 2011}.

\bibitem{settles2009}
B.~Settles, ``Active learning literature survey,'' University of
  Wisconsin--Madison, Computer Sciences Technical Report 1648, 2009.

\bibitem{srndic2013}
N.~{\v{S}}rndic and P.~Laskov, ``Detection of malicious {PDF} files based on
  hierarchical document structure,'' in \emph{NDSS}, 2013.

\bibitem{stringhini2013}
G.~Stringhini, C.~Kruegel, and G.~Vigna, ``Shady paths: Leveraging surfing
  crowds to detect malicious web pages,'' in \emph{ACM CCS}, 2013.

\bibitem{virustotal}
{VirusTotal}, \url{https://www.virustotal.com/}, retrieved on July 30, 2014.

\end{thebibliography}

\makeappendix

\end{document}